On the Influence of Artificial Intelligence on Human Problem-Solving: Empirical Insights for the Third Wave in a Multinational Longitudinal Pilot Study

**On the Influence of Artificial Intelligence on Human Problem-Solving: Empirical Insights for the Third Wave in a Multinational Longitudinal Pilot Study**

Matthias Huemmer 1*; Theophile Shyiramunda 2*; Franziska Durner 3*; Michelle J. Cummings-Koether 4*

**1* Prof. Dr.-Ing. Matthias Huemmer,**

Professor, Institute for the Transformation of Society (I-ETOS), Deggendorf Institute of Technology - European Campus Rottal-Inn, Germany

E-Mail: matthias.huemmer@th-deg.de

ORCID ID: https://orcid.org/0009-0003-8122-470X

**2* Dr. Theophile Shyiramunda,**

Postdoctoral Research Associate, Institute for the Transformation of Society (I-ETOS), Deggendorf Institute of Technology- European Campus Rottal-Inn, Germany

E-Mail: theophile.shyiramunda@th-deg.de

ORCID ID: https://orcid.org/0000-0001-6725-3756

**3* Franziska Durner, MA,**

Research Associate, Institute for the Transformation of Society (I-ETOS), Deggendorf Institute of Technology - European Campus Rottal-Inn, Germany

E-Mail: franziska.durner@th-deg.de

ORCID ID: https://orcid.org/0000-0002-8064-5364

**4* Prof. Dr. Michelle J. Cummings-Koether,**

Professor, Institute for the Transformation of Society (I-ETOS), Deggendorf Institute of Technology - European Campus Rottal-Inn, Germany

E-Mail: michelle.cummings-koether@th-deg.de

ORCID ID: https://orcid.org/0000-0002-7137-3539

**Corresponding author @ Prof. Dr.-Ing. Matthias Huemmer ,**

Professor, Institute for the Transformation of Society (I-ETOS), Deggendorf Institute of Technology - European Campus Rottal-Inn, Germany

E-Mail: matthias.huemmer@th-deg.de

ORCID ID: https://orcid.org/0009-0003-8122-470X

## Declarations

### Conflict of Interest

The authors have no relevant financial or non-financial interests to disclose.

### Funding

This research received no specific grant from any funding agency, commercial, or not-for-profit sectors.

### Authorship and Contributions

Michelle J. Cummings-Koether, and Matthias Huemmer led the conceptualization of the pilot study. The methodology for this report was developed by Matthias Huemmer and Theophile Shyiramunda. Data collection was carried out by Michelle J. Cummings-Koether, Matthias Huemmer and Franziska Durner, while formal analysis was performed by Matthias Huemmer and Theophile Shyiramunda. The original draft of the manuscript was prepared by Matthias Huemmer, with subsequent review and editing provided by Theophile Shyiramunda and Michelle J. Cummings-Koether and Matthias Huemmer. Supervision of the project was managed by Michelle Cummings-Koether and Matthias Huemmer, and project administration was undertaken by Matthias Huemmer, Michelle Cummings-Koether and Theophile Shyiramunda. All authors have read and approved the final manuscript.

### Use of Large Language Models (LLMs)

Use of AI assistance: An AI assistant (ChatGPT/GPT-5 Thinking) was used for language editing, outline refinement, and stylistic suggestions. All conceptual content, framework design, interpretations, and final decisions are the author's own. The AI system did not have authorship or decision-making roles and is not listed as an author.

### Ethics Approval and Consent to Participate

This study received ethics approval from [Institution/Ethics Board name] and was conducted in accordance with ethical guidelines for human subjects research and the General Data Protection Regulation (GDPR).

All participants provided informed consent prior to participating, in accordance with Article 6(1)(a) and Article 7 of the GDPR. Participants were informed of the study's purpose, procedures, potential risks and benefits, and their right to withdraw at any time without penalty. Consent was freely given, specific to this research, and documented through an explicit declaration.

Personal data were collected and processed lawfully on the basis of participant consent. All data were anonymized and securely stored in compliance with GDPR requirements to protect participant confidentiality and privacy throughout the research process and in perpetuity.

### Data Availability

All data and materials are available upon request.

**ABSTRACT**

This article is presenting the results and their discussion for the third wave (with n=23 participants) within a multinational longitudinal study that investigates the evolving paradigm of human-AI collaboration in problem-solving contexts. Building upon previous waves, our findings reveal the consolidation of a hybrid problem-solving culture characterized by strategic integration of AI tools within structured cognitive workflows. The data demonstrate near-universal AI adoption (95.7% with prior knowledge, 100% ChatGPT usage) primarily deployed through human-led sequences such as "Think, Internet, ChatGPT, Further Processing" (39.1%). However, this collaboration reveals a critical verification deficit that escalates with problem complexity. We empirically identify and quantify two systematic epistemic gaps: a belief-performance gap (up to +80.8 percentage points discrepancy between perceived and actual correctness) and a proof-belief gap (up to -16.8 percentage points between confidence and verification capability). These findings, derived from behavioral data and problem vignettes across complexity levels, indicate that the fundamental constraint on reliable AI-assisted work is solution validation rather than generation. The study concludes that educational and technological interventions must prioritize verification scaffolds (including assumption documentation protocols, adequacy criteria checklists, and triangulation procedures) to fortify the human role as critical validator in this new cognitive ecosystem.

## 1. Introduction

The pervasive integration of Artificial Intelligence (AI), particularly large language models (LLMs), into every day and professional contexts is catalyzing a fundamental transformation in how humans approach, reason through, and solve problems. These systems are reshaping the cognitive landscape, altering established patterns of information retrieval, analytical reasoning, and decision-making (Kasneci et al., 2023). This transformation finds a theoretical foundation in classic cognitive science, which posits that tools can become deeply integrated into a problem-solver's functional architecture, which is a concept known as the "extended mind" (Clark & Chalmers, 1998). From this perspective, AI acts not merely as a tool, but as an active participant in cognition, serving as both an external memory repository and a methodological resource that redistributes cognitive effort and fundamentally reconfigures problem-solving workflows (Hutchins, 1995).

A growing body of literature, including findings from the published Wave 1 of the I-ETOS study on AI's cultural impact (Huemmer et al., 2025a), demonstrates that digital tools reshape human memory and strategy. The phenomenon of "cognitive offloading" (Risko & Gilbert, 2016; Sparrow et al., 2011) suggests that when information is reliably accessible externally, individuals prioritize remembering how to access it over the information itself. While this can free up cognitive resources for higher-order reasoning, it also carries the risk of reduced knowledge internalization and the potential for what some scholars term "cognitive debt" if over-relied upon or miscalibrated (Kosmyna et al., 2025). The efficacy of this cognitive partnership hinges on calibrated trust. Decades of human-automation research illustrate that under-trust leads to the disuse of capable tools, while over-trust results in misuse and "automation bias" that is the uncritical acceptance of automated outputs without sufficient scrutiny (Lee & See, 2004; Parasuraman & Riley, 1997). These dynamics are critically relevant to LLM-assisted problem-solving, where outputs are often fluent and persuasive yet can be factually incorrect or logically flawed. It is a tendency known as "hallucination", refer to (Ji et al., 2023). This underscores the non-negotiable necessity of human verification, particularly as task complexity escalates (Huang et al., 2023).

Empirical evidence from field studies indicates that generative AI can enhance productivity and output quality on well-specified tasks, with notable gains among less-experienced workers (Brynjolfsson et al., 2025; Noy & Zhang, 2023). However, these benefits are highly task-dependent and may not generalize across all domains. Collectively, these findings frame AI as a powerful but double-edged aid to problem-solving, one that requires careful and critical integration into human workflows.

Against this theoretical and empirical backdrop, this article presents the findings from Wave 3 of our multinational longitudinal pilot study, conducted according to the methodology framework established by Huemmer et al. (2025c). Building directly upon Wave 1's exploration of AI's cultural impact (Cummings-Koether et al., 2025; Huemmer et al., 2025a;) and Wave 2's identification of critical verification deficits (Huemmer et al., 2025b), this third wave provides a crucial update on the evolving nature of human-AI collaboration. Wave 3 serves to consolidate our understanding of the hybrid problem-solving culture that is emerging and to provide a robust, quantitative assessment of the performance dynamics at play.

Our contribution is threefold. First, we provide updated descriptive results from a diverse sample (n=23), confirming the patterns of strategic AI integration and revealing its continued entrenchment in problem-solving workflows. Second, we deliver a definitive quantification of the two critical gaps in human-AI collaboration for the "belief-performance gap" (the discrepancy between perceived and actual correctness of AI solutions) and the "proof-belief gap" (the discrepancy between belief in correctness and the ability to verify it) using data from across the complexity spectrum. Third, we demonstrate the maturation of a hybrid problem-solving culture where AI is orchestrated within, rather than replacing, traditional cognitive processes. The findings from Wave 3 solidify the empirical foundation for understanding the evolving nature of human cognition in the age of AI and create an imperative for educational frameworks and system designs that foster critical evaluation skills to ensure reliable and effective collaborative outcomes.

## 2. Theoretical Framework: The AI-Human Problem-Solving Interaction Model

### 2.1. Integration of Human and AI Capabilities

The documented workflow sequences provide compelling empirical evidence for a hybrid intelligence model in which AI systems function as integrated cognitive tools rather than autonomous replacements for human expertise. Analysis reveals that the "Think, Internet, ChatGPT, Further Processing" pathway accounts for 39.1% of documented sequences (Table 5), demonstrating systematic human orchestration of AI within multi-stage workflows. This substantiates the theoretical claim that AI operates primarily as a complementary instrument augmenting human cognition rather than a substitutive technology displacing expert judgment (Davenport & Kirby, 2016; Huemmer et al., 2025b).

Sequential integration patterns, where initial human framing precedes AI consultation followed by verification and refinement, align precisely with the cognitive redistribution mechanism posited in the theoretical framework (Figure 1). Users demonstrably retain responsibility for problem decomposition, information triangulation across multiple sources including traditional internet search, output evaluation, and quality assurance. AI consultation occurs as an embedded step within this human-directed process rather than as an isolated activity, confirming that AI has been absorbed into existing problem-solving repertoires rather than displacing established methodological approaches.

**Figure 1.** The AI-Human Problem-Solving Interaction Framework.

**Source.** Authors' illustration based on the study findings in Huemmer et al. (2025b).

Figure 1 shows the AI-Human Problem-Solving Interaction Framework, synthesizing the study findings in Huemmer et al. (2025b) and combining it with cognitive load theory (Sweller, 1988) and the extended mind thesis (Clark & Chalmers, 1998) to conceptualize AI as an external cognitive resource. The framework illustrates how AI features, mediated by calibrated trust varying with problem complexity and domain context, trigger cognitive redistribution between human and machine processes. This enables users to offload solution generation while retaining critical verification preservation roles at key decision junctures. The framework operates recursively, where outcomes including efficiency gains and verification gaps subsequently reshape future interactions, competencies, and problem-solving identity (Floridi & Cowls, 2022; Glikson & Woolley, 2020; Huemmer et al., 2025b).

The verification-driven workflow in Figure 2 operationalizes these principles through a seven-stage process model supported by empirical sequences. The workflow begins with Problem Framing and Decomposition, where complex problems are analyzed and broken into manageable sub-components. Adequacy Criteria Definition establishes success metrics and thresholds before AI consultation to prevent post-hoc rationalization. Verification Planning specifies triangulation sources, replication methods, and documentation protocols. Resource Consultation strategically draws on domain knowledge, literature, tools, and AI for decomposed sub-tasks. Triangulation and Testing subjects AI outputs to independent checks using predefined tests. Documentation and Audit records assumptions, prompts, evidence, and decision lineage (Adadi & Berrada, 2018; Huemmer et al., 2025b). Decision and Delivery accept solutions only when adequacy criteria are met; otherwise, the process iterates.

Task-class overlays calibrate verification intensity by difficulty. Simple tasks require minimal single-check notes; difficult tasks demand dual-source triangulation and enhanced documentation; complex tasks necessitate replication across methods, sensitivity analysis, adversarial testing, and full audit trails. Dynamic recursion indicates that outcomes feed into future cycles by reshaping trust, competencies, and practice.

## 2.2. Verification Preservation and Task Complexity

The workflow data reveal persistent verification preservation behaviors consistent with the framework's prediction that users maintain human checkpoints at critical junctures (Huemmer et al., 2025b). Multi-stage sequences incorporating pre-AI preparation, AI consultation, and post-AI validation demonstrate calibrated delegation where users leverage AI for specific sub-tasks while retaining metacognitive oversight. The continued prominence of traditional resources including books, internet search, and colleague consultation provides triangulation points for AI output verification and maintains independent cognitive pathways preventing over-reliance on single-source responses (Hutchins, 1995).

Qualitative analysis reveals systematic variation in AI integration strategies across problem types, consistent with Figure 2's task-class overlay specification. For simple, well-defined problems, participants demonstrated direct consultation patterns with minimal elaboration, reflecting appropriate reduction of verification intensity. For complex, ambiguous challenges, workflows exhibited multiple information sources, iterative refinement cycles, and extended post-AI processing, suggesting adaptive calibration of verification effort to problem demands (Sweller, 1988; Zimmerman, 2000a). This pattern differentiation supports the proposition that effective AI collaboration requires context-sensitive trust calibration rather than uniform adoption or rejection (Venkatesh et al., 2003; Parasuraman & Riley, 1997).

However, instances where verification appeared being insufficient relative to task complexity, particularly under time pressure, highlight the verification gap vulnerability where AI output fluency may short-circuit necessary checking processes (Mosier et al., 1998; Huemmer et al., 2025b). This motivates structured competency development interventions proposed in subsequent research phases.

## 2.3. Competency Development and Learning Trajectories

The empirical confirmation of hybrid workflows validates the framework's emphasis on AI supervision skills as essential professional competencies requiring explicit cultivation (Vrontis et al., 2023). Observed patterns demonstrate four critical competency clusters. First, decomposition proficiency involves analyzing complex problems into coherent sub-components suitable for AI consultation. The prevalence of initial Think phases indicates user recognition of this need, though variability in effectiveness suggests explicit training is necessary. Second, strategic resource orchestration coordinates diverse information sources, leveraging AI for rapid generation, internet search for targeted retrieval, literature for authoritative knowledge, and colleagues for domain expertise (Ferrari, 2012; Huemmer et al., 2025b).

Third, verification discipline encompasses commitment and capability to validate AI outputs thoroughly. While post-AI processing persists in observed workflows, occasional insufficient checking indicates that verification self-efficacy represents a critical bottleneck requiring targeted development (Chen et al., 2024; Huemmer et al., 2025b). Fourth, documentation practices enable accountability and organizational learning. Variability in documentation extent highlights the need for organizational scaffolding through protocols, templates, and audit mechanisms that institutionalize verification beyond individual initiative (Adadi & Berrada, 2018; Singh et al., 2025).

Longitudinal comparison between Wave 1 (Huemmer et al., 2025b) and Wave 2 data suggests evolutionary adaptation in hybrid strategies, with increasing sophistication in verification approaches and task differentiation. This indicates practice-based learning as users accumulate experience with AI capabilities and limitations (Clark & Chalmers, 1998). However, some participants consolidated under-verified habits, suggesting experience alone does not guarantee appropriate calibration without deliberate reflection on outcomes (Zimmerman, 2000b). The observed formation of hybrid problem-solving identities, where AI consultation becomes normalized while verification values persist, represents significant cultural shift supporting sustainable AI integration (Floridi & Cowls, 2022; Glikson & Woolley, 2020).

## 2.4. Design and Organizational Implications

Empirical validation provides clear direction for system design and organizational practice. AI tools should support sequential, verification-rich workflows rather than optimizing solely for generation speed. Design priorities include verification scaffolds with prompts for adequacy criteria specification and structured checking workflows (Tankelevitch et al., 2024; Singh et al., 2025), multi-turn refinement support preserving conversational context for iterative improvement (Song, 2025), provenance transparency indicating information sources and uncertainty levels facilitating informed verification (Ryan, 2010; Chen et al., 2024), and automatic audit trails supporting documentation requirements without manual burden.

Organizations must complement technological capabilities with institutional supports. Training programs should develop AI supervision skills through explicit instruction in decomposition, triangulation, verification, and documentation. Protocols should calibrate verification intensity to task complexity through formal classification systems. Quality assurance mechanisms should detect verification gaps through audit sampling and outcome tracking. Cultural initiatives should legitimize time investment in thorough checking through leadership messaging and performance evaluation criteria.

The documented dominance of hybrid workflows offers an optimistic counterpoint to concerns about deskilling or uncritical automation dependence. Users maintain verification responsibilities and continue employing traditional information sources alongside AI tools, suggesting resilience of professional judgment. However, realizing this augmentation potential requires proactive cultivation of competencies, practices, and organizational conditions enabling reliable human-AI collaboration. The verification-centered framework and supporting empirical patterns provide actionable guidance for this critical developmental agenda, specifying both technical designs and human competency investments necessary for responsible AI integration in professional problem-solving contexts.

## 3. Materials and Methods

In its third wave (Wave 3), the longitudinal survey continues the investigation of AI, following a framework based on Huemmer et al. (2025a) methodological paper, grounded in the UTAUT model (Venkatesh et al., 2012), ensuring transparency, reproducibility, and consistency, and aligning with established survey research standards (Plano Clark, 2017) (see Figure 2). The overarching research design, including the theory-grounded framework, operationalization of constructs, and longitudinal procedure, is detailed in the companion methodological publication (Huemmer et al., 2025a). This framework prioritizes procedural transparency and reproducible reporting in line with established survey research guidelines (Plano Clark, 2017). The constructs, particularly those measuring AI adoption and perceptions, are contextualized within established technology-acceptance theories, such as the Unified Theory of Acceptance and Use of Technology (UTAUT) (Venkatesh et al., 2012), to enhance interpretability and scientific validity. Results from Wave 1 and Wave 2 of this longitudinal study, focusing on AI's intersection with culture and problem-solving, are published in Huemmer et al. (2025b), Huemmer et al. (2025c), and Cummings-Koether et al. (2025).

For Wave 3 a convenience sample of 23 participants was recruited from the European Campus Rottal-Inn community. All participants provided informed consent, and the ethical and data privacy procedures adhered to GDPR principles, remaining consistent with the protocol described in Huemmer et al. (2025a). The survey instrument consisted of structured questions grouped into key domains as outlined in Huemmer et al. (2025a) and applied consistently in Waves 1 and 2 as well as in this study (Huemmer et al., 2025b, c). These domains included demographics and role context, AI knowledge and usage patterns, perceived efficiency, reliability, safety, and ethics, problem-solving with AI, self-assessment of problem-solving methodology, and process sequences and vignettes. It should be noted that other aspects concerning the influence on human culture will be covered in a separate article and are not considered here.

Data processing and analysis were descriptive, focusing on frequency counts, percentages, and measures of central tendency (means, medians) and dispersion (standard deviations, interquartile ranges) for the entire sample. As a pilot study, no inferential tests were performed, and subgroup patterns are discussed qualitatively. The data handling procedures, from processing to reporting, followed the same reproducible procedures outlined in Huemmer et al. (2025a) and specifically as applied in Waves 1 and 2, which are also applied here for Wave 3 in Huemmer et al. (2025b, c).

Figure 2 below represents the overview of the methodology used for Study Wave.

**Figure 2.** Overview of Materials and Methods for Study Wave 3

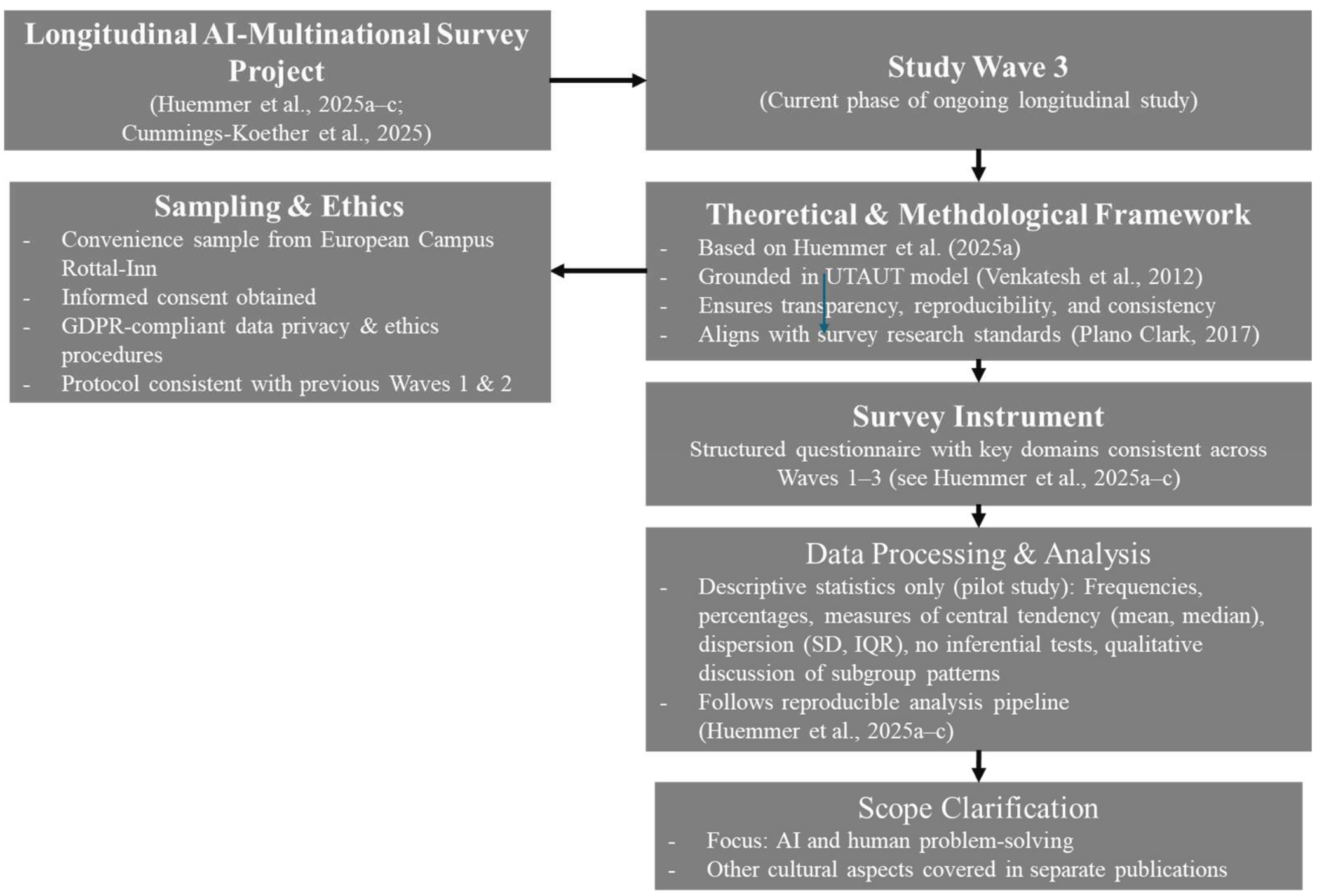

**Source**: Authors' own illustration based on the longitudinal survey framework (Huemmer et al., 2025a–c).

Figure 2 summarizes the methodological structure of Study Wave 3 within the longitudinal AI–Culture Survey Project. Building on the established framework of earlier waves (Huemmer et al., 2025a–c) and grounded in the UTAUT model, this phase integrates ethical and GDPR-compliant sampling, a structured survey instrument, and descriptive analytic procedures emphasizing transparency and reproducibility. The focus of this wave is exploratory, examining AI's role in human problem-solving, with broader cultural aspects discussed in companion studies.

Data processing and analysis were descriptive, focusing on frequency counts, percentages, and measures of central tendency (means, medians) and dispersion (standard deviations, interquartile ranges) for the entire sample. As a pilot study, no inferential tests were performed, and subgroup patterns are discussed qualitatively. The data handling procedures, from processing to reporting, followed the same reproducible procedures outlined in Huemmer et al. (2025a) and specifically as applied in Waves 1 and 2, which are also applied here for Wave 3 in Huemmer et al. (2025b, c).

## 4. Results

This section reports the descriptive results from Wave 3 (n = 23). All statistics are purely descriptive (i.e. counts, percentages, mean, SD, median, and IQR) and were calculated directly from the Wave 3 CSV dataset. Because several items were optional, we provide the per-item N for each result. All figures are generated from the same CSV to ensure traceability.

### 4.1. Participant Demographics

The Study Wave 3 sample (n = 23) was culturally diverse across four continents. By nationality, 26.1% were German, 39.1% were grouped as African and Asian (Egyptian, Ghanaian, Iranian, Nepali, Myanmar, Pakistani), 21.7% were from the Americas (Chilean, Ecuadorian, Guatemalan, Salvadorean, United States), and 13.0% did not report a nationality. Women constituted 60.9% of participants and men 39.1%, with no responses in the diverse category.

Age was balanced between Youth (<30 years; 52.2%) and Adults (30–69 years; 47.8%); no participants were ≥70 years. Educational attainment was high, where 52.2% held postgraduate qualifications (Master's/PhD/Habilitation) and 47.8% held non-postgraduate credentials (high school or bachelor's). Students formed 60.9% of the sample, with 39.1% non-students (teaching, administrative, or non-ECRI affiliates).

The overall demographics of the participants in Study Wave 3 are summarized in Table 1.

**Table 1.** Participants demographics.

| Characteristic | Category | Frequency (f) | Percentages % |
|---|---|---|---|
| **Age** | Youth (<30 years) | 12 | 52.2 |
| | Adults (30–69 years) | 11 | 47.8 |
| | Older adults (≥70 years) | 0 | 0 |
| **Gender** | Men | 9 | 39.1 |
| | Women | 14 | 60.9 |
| | Diverse | 0 | 0 |
| **Nationality** | German | 6 | 26.1 |
| | Africa (Egyptian, Ghanaian) | 3 | 13.0 |
| | Asia (Iranian, Nepali, Myanmar, Pakistani) | 6 | 26.1 |
| | Americas (Chilean, Ecuadorian, Guatemalan, Salvadorean, United States) | 5 | 21.7 |
| | No Answer | 3 | 13.0 |
| **Occupation** | Non-students (Teaching + Admin + NonECRI affiliation) | 9 | 39.1 |
| | Students | 14 | 60.9 |
| **Education Level** | Non-postgraduate (High school + Bachelor) | 11 | 47.8 |
| | Postgraduate (Master’s + PhD + Habilitation) | 12 | 52.2 |
| | Other (Dipl. Ing. FH) | 0 | 0.0 |

**Note.** Percentages are based on the total sample size (n = 23). Nationality categories were grouped for clarity (Europe, Africa, Asia, North America, South America); counts in parentheses show raw frequencies.

**Source.** Authors’ own illustration based on survey data.

### 4.2. AI Knowledge and Usage Patterns

As summarized in Table 2, participants reported high prior exposure to AI (95.7%), indicating a sample with substantial baseline familiarity. ChatGPT usage was universal (100%). Beyond this anchor tool, adoption was diversified but fragmented with participant using Google Gemini (34.8%), Perplexity (30.4%), Canva (30.4%), and Microsoft Copilot (26.1%) formed a second tier, while Claude (21.7%), MetaAI (17.4%), and Grok (13.0%) showed more selective uptake; Jasper was not used (0%). “Other” tools were collectively reported by 30.4% and included Deepseek (n = 4), NotebookLM, Napkin, Ideogram, Miro, Elicit, ResearchRabbit, Manus, and Gamma. This suggests an exploratory long-tail pattern. (Multiple responses were possible for tools).

**Table 2.** AI Knowledge and Usage Patterns.

| Characteristic | Category | Frequency (f) | Percentages % |
|---|---|---|---|
| **AI Knowledge** | Prior knowledge in AI | 22 | 95.7 |
| | No prior knowledge | 1 | 4.3 |
| **AI Tools Used** | ChatGPT | 23 | 100.0 |
| | Claude | 5 | 21.7 |
| | MetaAI | 4 | 17.4 |
| | Grok | 3 | 13.0 |
| | Google Gemini | 8 | 34.8 |
| | MS Copilot | 6 | 26.1 |
| | Perplexity | 7 | 30.4 |
| | Jasper | 0 | 0% |
| | Canva | 7 | 30.4 |
| | Others (Deepseek [4 times mentioned], Manus NotebookLM, Napkin, ideogram, miro, elicit, Researchrabitt, Gamma): | 7 | 30.4 |
| **Contexts of AI Use** | Private | 18 | 78.3 |
| | Professional | 21 | 91.3 |
| | Educational | 19 | 82.6 |
| **Frequency of AI Use** | Several times a day and once a day | 22 | 95.7 |
| | Several times a week and once a week | 0 | 0.0 |
| | Other | 1 | 4.3 |

**Note.** Percentages are based on the total sample size (n=23). Multiple responses were possible for AI tools.

**Source.** Authors' own illustration based on survey data.

As shown in Table 2, AI was integrated most strongly in professional contexts (91.3%), followed by educational (82.6%) and private use (78.3%), indicating broad but work-led adoption. In terms of cadence, daily engagement predominated. Here, 95.7% reported using AI several times a day or once a day; no respondent reported weekly use, and 4.3% selected "other." This pattern is consistent with routinized incorporation of AI into everyday workflows rather than occasional or exploratory use.

Collectively, the data depicts a pervasively adopted, ChatGPT-centric ecosystem with selective multi-tool supplementation and highest penetration in professional tasks. While the small sample (n = 23) and optional responses warrant caution, the consistent daily usage across respondents supports the interpretation that, for this cohort, AI tools have transitioned from ad-hoc assistance to standard operating resources in professional, educational, and secondarily private domains.

### 4.3. Perceptions of AI

Perceptions of AI were moderate and context-sensitive (Table 3). On reliability, respondents reported mid-scale confidence (mean = 4.9 on a scale from 1 = not reliable to 8 = reliable, SD = 1.0), indicating AI is regarded as neither clearly unreliable nor highly trustworthy. Data protection concerns were likewise moderate (mean = 4.9 on a scale from 1 = *protected* to 8 = *unprotected/unsafe,* SD = 1.7), suggesting a slight tilt toward caution but not pronounced alarm.

Ethical evaluations exhibited the strongest domain effects. A majority judged AI use not to be cheating in professional (69.6%) and private (65.2%) contexts, whereas use in academic theses was considered cheating by 69.6% of respondents. Taken together, the pattern reflects a calibrated ethical stance that combines permissive norms in everyday and workplace applications with stricter academic integrity expectations, even as overall reliability and safety perceptions remain moderate rather than polarized.

**Table 3.** Perceptions of AI (n=23).

| Dimension | Findings | % / M (SD) |
|---|---|---|
| **Perceived Reliability** | AI considered moderately reliable (mean = 4.1 on 1–8 scale) | M = 4.9 (±1.0) |
| **Data Safety** | Concerns about data protection; many felt unsafe (mean = 5.5 on 1–8 scale, where 1 = protected, 8 = unsafe) | M = 4.9 (±1.7) |
| **Ethical Use** | Use in private contexts not widely considered problematic | 65.2% said "not cheating" |
| | Use in professional contexts not widely considered problematic | 69.6% said "not cheating" |
| | Use in academic theses often seen as problematic (plagiarism risk) | 69.6% said "cheating" |

**Note.** Reliability and safety were measured on 8-point Likert scales (1 = low, 8 = high). Percentages reflect categorical survey responses.

**Source.** Authors' own illustration based on survey data.

### 4.4. Anticipated Influence of AI on Work and Private Life

Participants expect AI's influence to increase across domains over the next 2–5 years (Table 4). For working behavior, the mean rises from mean = 5.0 (SD = 2.0) at present to mean = 5.9 (SD = 2.1), a +0.9 shift on the 8-point scale. For private behavior, the mean value increases from mean = 4.1 (SD = 2.3) to mean = 5.1 (SD = 2.1), a +1.0 shift. The current work > private gap persists into the future (work remains higher than private), indicating stronger (and increasingly work-led) integration.

The relatively large SDs (≈2.0 to 2.3) signal substantial heterogeneity, where some participants already experience high impact, while others report low or moderate influence. Nonetheless, the parallel upward shifts in both domains suggest a broad trajectory toward deeper routine integration, with the workplace poised to remain the primary locus of near-term transformation.

**Table 4.** Perceived influence of AI on working and private behavior (present vs. near future).

| Behavior | Timeframe | Influence Score M (SD) |
|---|---|---|
| Working behavior | Present | M = 5.0 (±2.0) |
| *(where 1 = low implications and 8 = high implications)* | Future | M = 5.9 (±2.1) |
| Private behavior | Present | M = 4.1 (±2.3) |
| *(where 1 = low implications and 8 = high implications)* | Future | M = 5.1 (±2.1) |

**Source.** Authors' own illustration based on survey data.

## 4.5. Problem-Solving Methodologies and AI Integration

As detailed in Table 5, the participants reported hybrid, multi-step workflows for difficult/complex problems, with AI typically consulted after an initial human ideation phase. The most frequent sequence was "Think, Internet, ChatGPT, Further Processing" (39.1%), indicating a human-led → broad search → AI synthesis pipeline rather than immediate AI reliance. A traditional non-AI pathway ("Think, Paper, Sketch, Book, Further Processing") remained substantial (21.7%), evidencing continued use of analog/expert sources alongside digital tools.

**Table 5.** Proceeding to solve a difficult or complex problem (n=23).

| Proceeding for complex problem-solving | Percentage |
|---|---|
| Think, Paper, Sketch, Book, Further Processing | 21.7% |
| Think, ChatGPT, Further Processing | 8.7% |
| Think, Internet, Further Processing | 0.0% |
| Think, Internet, ChatGPT, Further Processing | 39.1% |
| ChatGPT, Think, Further Processing | 4.3% |
| ChatGPT, Internet, Think, Further Processing | 13.0% |
| ChatGPT, Reformulation, ChatGPT, Reformulation | 0.0% |
| Other: Please specify your answer | 13.0% |

**Note.** Other category responses are given in Appendix Table 15.

**Source.** Authors' own illustration based on survey data.

Data in Table 5 shows that AI-initiated paths were less common overall as it is shown by "ChatGPT, Think, Further Processing" (4.3%) and "ChatGPT, Internet, Think, Further Processing" (13.0%) together accounted for 17.3%. Notably, "Think, Internet, Further Processing" without AI was 0.0%, suggesting that once web search is engaged, many respondents subsequently bring ChatGPT into the loop for integration or verification. Pure iterative prompting without external checks ("ChatGPT, Reformulation, ChatGPT, Reformulation") was also 0.0%, implying limited reliance on closed AI-only refinement cycles.

The "Other" category (13.0%) underscores verification-centric iteration, e.g., "Think → Paper → ChatGPT → Further Processing → Reformulation → ChatGPT → Double-check with Book/Internet." These free-text sequences reveal alternation between AI outputs and authoritative sources, consistent with metacognitive control and quality assurance.

These findings in Table 5 for Wave-3 respondents show that they predominantly use human-led, hybrid strategies where AI is a mid-pipeline collaborator (post-ideation, often post-search) rather than a first touch. Traditional methods persist, AI-only loops are rare, and verification with external sources is a salient feature of mature problem-solving practice.

**Analysis of Problem-Solving Strategies Across Complexity Levels**

An analysis of the Analysis of Problem-Solving Strategies of the participants (detailed in Appendix Table 16, Table 17 and Table 18) show that he responses depict a clear progression from human-led initiation toward increasingly hybrid and iterative workflows as task demands rise. The pattern is not uniform, since a few participants introduce AI at very early stages even for simple or difficult tasks and one respondent indicates having no complex problems at all. Overall, however, the center of gravity moves from self-reliant thinking and quick execution to structured sequences that combine multiple sources, reformulation, and verification.

For simple problems (see Appendix Table 16), most respondents begin with autonomous cognition and light scaffolding. Typical descriptions include think and solve, think and calculate, think and sketch, or think and further processing. External aids are introduced as needed rather than by default. Several participants describe first attempting a solution using their own knowledge, then consulting a book to refresh specifics or searching the internet if memory fails. ChatGPT appears frequently as a secondary step for verification or for moving past a sticking point, captured in sequences such as think, further processing, ChatGPT and think, internet, ChatGPT, further processing. Paper-based visualization and brief written organization of ideas remain common, suggesting that analog supports are still valued for clarity at low complexity.

As tasks become difficult (see Appendix Table 17), the approach grows more systematic and multi-source, while still often starting with an individual thinking or sketching step. Respondents report on layering internet research, YouTube videos, books, and AI consultation for analysis, verification, and reformulation. Examples include think, internet, ChatGPT, book, asking others and think, calculate or sketch with ChatGPT only if needed. There are also instances of AI being engaged earlier, such as ask ChatGPT or think, AI, further processing, indicating that some participants view AI as a first line of analysis once a problem is perceived as nontrivial. References to reformulation and checks against Google or other resources show an emphasis on triangulation and quality control rather than single source reliance.

When problems are described as complex (see Appendix Table 18), workflows expand into comprehensive and recursive pipelines that integrate library or internet research, books, academic papers, videos or courses, and multiple interactions with AI, often with explicit reformulation and checking. Participants describe research in internet or library, thinking, pairing results with generative AI, and reflection, as well as cycles like think, research, ChatGPT, check, reformulation if required, ChatGPT, check. Some sequences start with think, internet, AI, reform or think, internet, reformulation, which underscores the presence of iterative control loops aimed at refining problem understanding and solution quality. One respondent notes not having complex problems, which points to heterogeneity in perceived task complexity within the cohort.

Taken together, the data supports a hybrid problem solving culture that prioritizes an initial individual cognition phase and then layers external resources as complexity increases. AI is used selectively for simple tasks, becomes a collaborative partner for analysis, verification, and refinement in difficult tasks, and is integrated as part of a multi-source, iterative process in complex scenarios. The presence of AI first starts for some respondents and AI involvement, even in some simple cases indicates flexibility and adaptation rather than a strictly linear progression. Verification through cross checking with books, internet sources, and reformulation steps is a recurring feature of mature workflows across the spectrum.

**Means applied for problem-solving**

The vast majority of participants (95.7%) reported having knowledge of or using AI, confirming AI's broad diffusion into daily routines. An analysis of the general problem-solving methodologies employed (Table 6) reveals that consulting the Internet (95.7%) is a near-universal practice. This is complemented by other methods such as Asking others (78.3%), using Books (69.6%), and using a Mobile phone (60.9%). This suggests that AI is integrated into a diverse toolkit, augmenting rather than replacing established methods, particularly direct information retrieval and social consultation.

**Table 6.** Means applied for problem-solving (n=23).

| Methodology | Percentage |
|---|---|
| Books | 69.6% |
| Internet | 95.7% |
| Mobile phone | 60.9% |
| Library | 47.8% |
| Asking others | 78.3% |
| Guessing/Estimation | 43.5% |
| Other: | 13.0% |

**Source.** Authors' own illustration based on survey data.

**Functional Domains and Workflow Stages of AI Deployment**

The specific phases and areas, where AI is applied are detailed, in Table 7. AI use is most prominent in research (82.6%) and checking correctness (69.6%), underscoring its core role as a tool for information gathering and validation. It is also heavily integrated into regular works (65.2%), indicating its utility for everyday tasks. In contrast, uptake is considerably lower for visual and creative tasks like painting (8.7%) and sketching (13.0%), pointing to a task-dependent adoption pattern where AI is currently favored for analytical, text-based, and research-oriented tasks. The "Other" category (13.0%), which includes applications like using AI as a "Socratic dialog partner," further highlights its role in facilitating complex thought processes.

**Table 7.** Phases and areas where the participants apply AI (n=23).

| Areas where Participants apply AI | Percentage |
|---|---|
| Ideation | 47.8% |
| Creative works | 30.4% |
| Regular works | 65.2% |
| Formulations | 43.5% |
| Painting | 8.7% |
| Sketching | 13.0% |
| Programming | 34.8% |
| Checking of correctness | 69.6% |
| Research | 82.6% |
| Other | 13.0% |

**Note.** Other category responses that were given are "*Language clarity*", *"Learning"* and *"socratic dialog partner*".

**Source.** Authors' own illustration based on survey data.

**AI usage depending on problem categories**

Furthermore, as shown in Table 8, AI is applied across all problem categories. Contrary to initial assumptions, usage is highest for difficult tasks (73.9%), while it remains substantial for simple (56.5%) and complex (47.8%) problems. This pattern reflects AI's perceived utility as a versatile tool, with its primary role being that of a collaborative partner for tackling challenging, yet well-defined, problems rather than just routine or highly complex scenarios.

**Table 8.** Usage of AI across problem categories (n=23).

| Problem Category | Yes, I use AI |
|---|---|
| Simple | 56.5% |
| Difficult | 73.9% |
| Complex | 47.8% |
| Does not apply | 4.3% |

**Source.** Authors' own illustration based on survey data.

### 4.6. Task-Specific Problem-Solving: AI Consultation and Performance

#### 4.6.1. Problem-Solving Vignettes: Design and Rationale

As shown in Huemmer et al. (2025a), the survey employed four problem vignettes arranged along a graduated complexity continuum to probe how AI use varies with cognitive demand, following established sequenced-task paradigms for studying cognitive offloading (Risko & Gilbert, 2016). The same vignette set was administered in Wave 1 (Huemmer et al., 2025b) and Wave 2 (Huemmer et al., 2025c). Problem 1 (Simple) was "What is 20% of 150?". It operationalizes a baseline arithmetic task to test thresholds of "cognitive convenience" and effort-conservation usage (Gerlich, 2024, 2025), amid evidence that even elementary calculations are offloaded to digital tools (Shanmugasundaram & Tamilarasu, 2023). Problem 2 (Difficult) regarded a question about a triangle right-angle identification via the Pythagorean theorem. It allows us to assess whether AI is invoked for verification versus solution generation when declarative and procedural knowledge are required; it also probes depth-effort trade-offs (Stadler et al., 2024) and the expertise-reversal effect in human-AI collaboration (Chen et al., 2017). Problem 3 (Complex) was a two-draw dependent-probability item. It targets multi-step combinatorial reasoning that elevates intrinsic load (Sweller, 1988), thereby testing AI-dependency in chained reasoning and confidence-verification mismatches (Zhang et al., 2024; Lee et al., 2025), including risks of metacognitive laxity under assistance (Fan et al., 2024). Problem 4 (Complex/Linear Programming) utilized a constrained profit-maximization model, which captures the apex of modeling complexity where constraint satisfaction, counterintuitive optima, and prompt specificity shape outcomes; it thus interrogates known AI weaknesses in constraint-based reasoning and the need for prompt optimization (Chen et al., 2024; Shojaee et al., 2025). Sequencing was informed by Cognitive Load Theory and recent AI-assisted problem-solving research (Kasneci et al., 2023; Sweller, 1988), enabling analysis of how computational intensity, conceptual depth, stepwise integration, and modeling demands systematically modulate AI engagement, self-efficacy, and verification behavior across the problem-solving workflow.

#### 4.6.2. Self-assessment on correctness of AI-generated results per Problem Category

Table 9 summarizes participants' judgments of AI output accuracy for simple, difficult, and complex problems. For simple items, confidence is high, with 47.8% judging all results correct and 30.4% reporting that the majority were correct, while 21.7% indicated only some were correct and none selected that no results were correct. For difficult items, confidence remains comparatively strong but more distributed. Exactly 36.4% selected all correct and 36.4% majority correct, with 27.3% choosing only some correct and no responses indicating none correct. For complex items, responses shift toward a more cautious stance. Most of the respondents (54.5%) indicated only some results were correct, 40.9% reported the majority were correct, 4.5% selected all correct, and none selected that no results were correct.

**Table 9.** Participants belief about Answer correctness of AI-generated results per Problem Category.

| Correctness of AI-generated results | Problem Category | | |
|---|---|---|---|
| | Simple | Difficult | Complex |
| All results are correct | 47.8% | 36.4% | 4.5% |
| Some results are correct | 21.7% | 27.3% | 54.5% |
| Majority of results are correct | 30.4% | 36.4% | 40.9% |
| No results are correct | 0.0% | 0.0% | 0.0% |
| No Answers | 0.0% | 0.0% | 0.0% |
| | n=23 | n=22 | n=22 |

**Source.** Authors' own illustration based on survey data.

Taken together, the data in Table 9 indicates a calibrated pattern rather than a collapse of confidence as complexity rises. Simple problems elicit strong accuracy expectations. Difficult problems still show substantial confidence, with a near even split between all correct and majority correct. Complex problems prompt greater caution, reflected in the plurality selecting only some correct, yet the continued presence of majority correct, and the absence of any no correct responses suggest that participants generally expect mixed but usable performance and adjust their verification practices accordingly.

#### 4.6.3. Participant Self-Assessment of Validation Capability and AI Output Verification

Participants reported progressively lower confidence as problem complexity increased, and this is visible in both determination and verification rates, see Table 10,. For simple problems, 100.0% stated they could determine the solution on their own, while 86.4% felt able to verify its correctness. For difficult problems, these figures dropped to 72.7% for determination and 68.2% for verification. For complex problems, confidence was lowest, with 63.6% able to determine a solution and 59.1% able to verify it.

**Table 10.** Self-Assessed Capacity for Independent Determination and Verification of AI Output Accuracy.

| Problem Category | Agreement (Yes) Can Determine on Own | | Agreement (Yes) Can Verify and Check Correctness | |
|---|---|---|---|---|
| Simple | 100.0% | n=23 | 86.4% | n=22 |
| Difficult | 72.7% | n=22 | 68.2% | n=22 |
| Complex | 63.6% | n=22 | 59.1% | n=22 |

**Source.** Authors' own illustration based on survey data.

As seen in Table 10, the gap between determining and verifying is largest for simple items (13.6 percentage points, 100.0% vs. 86.4%) and narrows at higher complexity (4.5 points for difficult and 4.5 points for complex). This pattern suggests that, even when participants can produce an answer, they recognize additional checking is needed, and that increasing ambiguity and variable interactions constrain both solution generation and confident validation. Per-item sample sizes differ because some questions were optional.

The relationship between perceived difficulty and the decision to consult AI varies by problem type rather than following a simple monotonic rule, as shown in Table 11. For the percentage problem in Item 1, every respondent classified it as simple (100.0 percent, n = 23) and very few consulted ChatGPT (4.3 percent yes, 95.7 percent no). The geometry problem in Item 2 shows a mixed difficulty profile with 39.1 percent simple, 56.5 percent difficult, and 4.3 percent complex, and it also exhibits notable indecision about consulting AI, with 39.1 percent yes, 39.1 percent no, and 21.7 percent undecided. The probability problem in Item 3 has the highest share perceived as complex within that item (43.5 percent), and it also records the highest consultation rate in Table 11 at 69.6 percent. The optimization problem in Item 4 carries the highest overall complexity rating across items (60.9 percent complex) but its consultation rate is 52.2 percent, which is lower than the probability problem's consultation rate. Across all items, no one selected not solvable, indicating respondents felt able to attempt each problem. These patterns suggest that domain familiarity, perceived verifiability, and workflow habits complement perceived complexity when respondents decide whether to consult AI.

**Table 11.** Perceived difficulty of problems and subsequent usage of AI for solving them.

| Problem | Simple | Difficult | Complex | Not Solvable | n | Ask ChatGPT (Yes) | Ask ChatGPT (No) | Ask ChatGPT (Undecided) | n |
|---|---|---|---|---|---|---|---|---|---|
| 1 | 100% | 0.0% | 0.0% | 0.0% | 23 | 4.3% | 95.7% | 0.0% | 23 |
| 2 | 39.1% | 56.5% | 4.3% | 0.0% | 23 | 39.1% | 39.1% | 21.7% | 23 |
| 3 | 17.4% | 39.1% | 43.5% | 0.0% | 23 | 69.6% | 30.4% | 0.0% | 23 |
| 4 | 17.4% | 21.7% | 60.9% | 0.0% | 23 | 52.2% | 47.8% | 0.0% | 23 |

**Source.** Authors' own illustration based on survey data.

Finally, for the subset of attempts where ChatGPT was consulted (see Table 12), perceived correctness is consistently high across items, while the ability to prove correctness is lower in every case. In Item 1, 100.0 percent of those who consulted ChatGPT believed the answer was correct and 90.9 percent reported they could prove it (n = 22 for this row in Table 12). In Item 2, 93.8 percent believed the answer was correct and 81.0 percent could prove it (n = 21). In Item 3, belief in correctness is 88.2 percent and the ability to prove drops to 71.4 percent (n = 21). In Item 4, belief remains high at 93.8 percent and the ability to prove is 80.0 percent (n = 22). The gap between thinking the answer is correct and being able to prove it ranges from 9.1 percentage points in Item 1 to 16.8 points in Item 3, which indicates a persistent verification shortfall that is most pronounced in the multi-step probability setting. The consultation rates reported in Table 12 differ slightly from Table 11 due to rounding and per item response counts, but the internal pattern remains the same, showing that trust in AI outputs is high when it is used, while verification lags and tends to weaken as problems require more intricate reasoning. Self-assessed ability to definitively prove that the solution was correct was lower for every problem and showed a decline as complexity increased. This gap between trust in AI's output and the ability to independently verify it underscores a potential reliability issue and a salient "trust but verify" challenge, particularly as tasks become more complex.

**Table 12.** AI Consultation Patterns and Self-Assessed Verification Capabilities for Specific Problems.

| Problem | Asked ChatGPT | Think that it is Correct | Can Prove that the Solution is Correct | N |
|---|---|---|---|---|
| 1 | 4.5% | 100.0% | 90.9% | 22 |
| 2 | 43.5% | 93.8% | 81.0% | 21 |
| 3 | 59.1% | 88.2% | 71.4% | 21 |
| 4 | 63.6% | 93.8% | 80.0% | 22 |

**Source.** Authors' own illustration based on survey data.

#### 4.6.4. Belief-Performance and Proof-Belief Gaps (Correct Answers per Problem Category)

Performance was coded per item as correct or false, revealing a strong inverse relationship between task complexity and accuracy, see Table 13. Simple arithmetic in Problem 1 yielded 20 correct out of 21 responses (95.2%). Geometry in Problem 2 remained high at 17/21 correct (81.0%). Accuracy declined for the probability item, Problem 3, with 12/18 correct (66.7%). Linear programming in Problem 4 showed the sharpest drop, with 11/23 correct (47.8%). Aggregated across items, respondents produced 60 correct answers out of 83 total attempts (72.3%). The notable gap between Problem 3 and Problem 4 suggests a significant escalation in cognitive demand associated with constraint formulation and optimization.

**Table 13.** Summary of Correct Answers by Problem.

| Problem | Total Responses (n) | Correct Answers | False Answers | Percentage Correct |
|---|---|---|---|---|
| Problem 1 | 21 | 20 | 1 | 95.2% |
| Problem 2 | 21 | 17 | 4 | 81.0% |
| Problem 3 | 18 | 12 | 6 | 66.7% |
| Problem 4 | 23 | 11 | 12 | 47.8% |
| Total | 83 | 60 | 23 | 72.3% |

**Source.** Authors' own illustration based on survey data.

As can be seen in Table 13, the linear programming item displays a distinctive and informative error pattern. A large portion of respondents converged on the same suboptimal plan of 80 units of Product A and 20 units of Product B, yielding a profit of 4,600, while only 47.8 percent produced the correct plan of 100 units of Product A and 0 units of Product B, which yields 5,000. This clustering suggests difficulty with interpreting and operationalizing the inequality constraint that production of Product B cannot exceed that of Product A by more than 20 units. Many appear to have treated the inequality as an equality to be satisfied exactly or adopted a diversification heuristic that keeps both products in the plan, rather than testing profit at extreme feasible points. Under the total production cap of 100 units and a higher unit profit for Product A than Product B, any substitution of B for A reduces profit by 20 per unit. Systematically exploring the feasible region by checking corner solutions would therefore reveal that 100 units of A and 0 of B is optimal and that the popular 80–20 choice sits on a feasible boundary but is not profit maximizing. The observed pattern points to gaps in translating verbal constraints into inequalities, in using corner testing or a graphical method, and in applying a simple marginal reasoning check that compares unit profits under a binding capacity constraint.

Variation in per-item response counts, which range from 18 to 23, indicates modest nonresponse that likely co-varies with perceived difficulty and time-on-task. The probability item has the lowest number of responses at 18, aligning with its higher perceived complexity and the relatively high consultation rate reported for that problem type. When these completion patterns are read alongside self-assessments, a consistent story emerges. Table 10 shows that the share of participants who believe they can verify a result drops from 86.4 percent for simple items to 59.1 percent for complex items, while Table 12 shows that, when ChatGPT is consulted, belief in correctness remains high across problems but the ability to prove correctness is lower in every case and declines most for the probability item. The objective results in Table 13, particularly the 47.8 percent accuracy on the optimization task, indicate a belief-performance gap that widens as tasks require formal constraint handling and systematic search. Future iterations of the study should therefore combine self-assessed capability with embedded verification prompts, require explicit statement of constraints and corner checks, and log the intermediate reasoning steps. These adjustments would help disentangle effects of time pressure from conceptual difficulty, improve comparability across items with different response rates, and target the specific skills that appear to limit performance on optimization and other high-structure problem types.

## 5. Discussion of Results: AI Problem-Solving and User Behavior

The empirical results from Study Wave 3 (n = 23) delineate an emergent paradigm of human-AI collaboration characterized by the strategic integration of artificial intelligence into structured, multi-stage cognitive workflows. This paradigm is defined by a systematic approach where AI is predominantly deployed as a secondary resource following initial human problem framing and information gathering, rather than as a primary solution mechanism. Critically, this integration reveals a significant and escalating verification deficit as problem complexity increases, manifesting in two analytically distinct but functionally interrelated cognitive gaps: Proceeding between confidence in solutions and the demonstrable capability to verify them). These findings align with and extend recent research on automation bias and epistemic dependence (Wang et al., 2020).

### 5.1. Calibrated AI adoption and the emergence of hybrid intelligence

The data from Wave 3 reveals a sophisticated and near-universal adoption pattern. The sample demonstrated substantial baseline familiarity with AI (95.7%), and usage was anchored by ChatGPT (100%), forming a pervasively adopted, ChatGPT-centric ecosystem (see Table 2). Beyond this, tool adoption was diversified but fragmented, with a second tier including Google Gemini (34.8%), Perplexity (30.4%), and Canva (30.4%), and a long tail of exploratory tools (30.4%), indicating selective multi-tool supplementation.

This adoption was strongly integrated into daily routines, with 95.7% of participants using AI daily and most prominently in professional (91.3%), educational (82.6%), and private (78.3%) contexts (Table 2). This pattern indicates a transition from ad-hoc assistance to a standard operating resource, with the workplace as the primary locus of integration, supporting the concept of routinized incorporation into everyday workflows.

Most critically, the documented workflow sequences, particularly the dominance of the "Think, Internet, ChatGPT, Further Processing" pathway (39.1%), substantiate a hybrid intelligence model (Table 5). These patterns indicate that AI systems operate primarily as augmentation tools rather than replacement technologies (Davenport & Kirby, 2016), with human cognition retaining centrality in problem framing, information synthesis, and quality assurance. This model confirms that AI systems primarily function as complementary tools within a broader cognitive toolkit that includes traditional methods like using books (69.6%) and asking others (78.3%) (Table 6). This represents a deliberate distribution of cognitive labor (Hutchins, 1995), where AI is a mid-pipeline collaborator for synthesis and verification rather than a first touchpoint, underscoring a human-led, scaffolded approach to problem-solving. This graded adoption strategy, where AI consultation increases with perceived task complexity (Table 11), provides empirical support for models emphasizing perceived usefulness and task-technology fit as determinants of technology utilization (Venkatesh et al., 2003).

### 5.2. The verification crisis: belief-performance and proof-belief gaps

A central and critical finding of Wave 3 is the precise quantification of where user confidence, verifiability, and objective outcomes diverge in AI-assisted problem-solving. As detailed in Table 14, the integrated metrics across the four problem vignettes reveal two diagnostic indices that expose a severe metacognitive miscalibration that intensifies with complexity.

**Table 14.** Integrated calibration across problems (Wave 3).

| Problem | Perceived Difficulty Profile | AI Consulted | Perceived Correctness (if AI used) | Provable Correctness (if AI used) | Actual Correctness | Belief-Performance Gap | Proof-Belief Gap |
|---|---|---|---|---|---|---|---|
| **1 (Simple)** | 100.0% Simple | 4.3% | 100.0% | 90.9% | 90.5% | +9.5 pp | -9.1 pp |
| **2 (Difficult)** | 39.1% Simple, 56.5% Difficult 4.3% Complex | 39.1% | 93.8% | 81.0% | 81.0% | +12.8 pp | -12.8 pp |
| **3 (Complex)** | 17.4% Simple, 39.1% Difficult, 43.5% Complex | 69.6% | 88.2% | 71.4% | 66.7% | +21.5 pp | -16.8 pp |
| **4 (Complex)** | 17.4% Simple, 21.7% Difficult, 60.9% Complex | 52.2% | 93.8% | 80.0% | 13.0% | +80.8 pp | -13.8 pp |

**Note.** Perceived and provable correctness are calculated among respondents who consulted ChatGPT for the respective item; actual correctness reflects all respondents who attempted the item. Gap calculations are performed via: Belief-Performance = Perceived Correctness - Actual Correctness; Proof-Belief = Provable Correctness - Perceived Correctness. Data derived from Table 11, Table 12 and Table 13.

**Source.** Authors' own illustration based on survey data.

The observed patterns in Table 14 reveal three critical, interrelated phenomena. First, an inverse relationship exists between problem complexity and objective performance (90.5% correct for Problem 1 declining to 81.0% for Problem 2, then 66.7% for Problem 3, and finally 13.0% for Problem 4), while AI consultation rates generally increase from 4.3% to 39.1% to 69.6%, before declining slightly to 52.2% for Problem 4 (Table 11). This indicates that while users generally turn to AI for more challenging tasks, they are doing so precisely in the domains where the technology's reliability is most variable and where their own performance plummets, creating a potential dependency cycle for complex tasks.

Second, and more profoundly, the data exposes a double calibration problem. The substantial positive belief-performance gaps, which escalate from +9.5 percentage points for Problem 1 to +12.8 points for Problem 2, then +21.5 points for Problem 3, demonstrate that users develop significant overconfidence in AI-generated solutions that are divorced from reality. This is most starkly visible in Problem 4, where 93.8% of users who consulted AI trusted the output, yet only 13.0% of all respondents who attempted the problem provided the correct answer, yielding a catastrophic gap of +80.8 percentage points (see Table 14). This finding extends previous research on the illusion of explanatory depth (Rozenblit & Keil, 2002) by demonstrating how AI-generated outputs can exacerbate overconfidence through fluent but potentially misleading explanations (Ji et al., 2023).

Simultaneously, the pronounced negative proof-belief gaps reveal a more fundamental crisis, as users know they cannot verify what they nonetheless believe to be correct (Table 14). These gaps remain relatively stable across problems, ranging from -9.1 percentage points for Problem 1 to -12.8 points for Problem 2 and -16.8 points for Problem 3, with a slight reduction to -13.8 points for Problem 4. This pattern underscores that the fluent outputs of LLMs create a powerful illusion of understanding that bypasses users' natural skepticism, making them confident in answers they cannot personally validate (Singh et al., 2025). This is a direct consequence of the cognitive burden shifting from solution generation to solution validation, a burden that often exceeds the user's available mental resources or domain expertise as complexity rises. The collapsing self-assessed ability to verify results, from 86.4% for simple problems to 59.1% for complex problems (Table 10), confirms this verification paradox.

## 5.3. Cognitive Architecture of AI-Assisted Problem-Solving and the Verification Paradox

The documented workflow sequences from Wave 3 both confirm and refine our understanding of the cognitive architecture underlying AI-assisted problem-solving. The data strongly confirms the proceeding of scaffolded cognition first identified in earlier waves, where AI is deployed as an external resource *after* initial problem representation (Risko & Gilbert, 2016). This is empirically demonstrated by the continued dominance of the "Think, Internet, ChatGPT, Further Processing" pathway (39.1%, see Table 5), which now emerges as the canonical hybrid workflow.

However, Wave 3 reveals a critical new pattern in the near-total disappearance of the non-AI digital search. The "*Think, Internet, Further Processing*" pathway was reported by 0.0% of participants (Table 5). This suggests a significant shift in digital literacy norms, where consulting the Internet and consulting an AI tool are no longer distinct steps but are becoming functionally fused. The Internet search now primarily serves as a precursor to AI synthesis, rather than a terminal solution strategy.

This refined architecture intensifies the verification paradox. The cognitive burden is not merely shifting from solution generation to validation; it is being compounded by the need to validate the output of a system (the AI) that has itself processed information from another complex, unvetted system (the Internet). This paradox is evidenced by the collapsing metrics for complex tasks, where AI consultation for Problem 3 was high (69.6%, Table 11), the actual correctness was only 66.7%, and the ability to prove that correctness was just 71.4% (Table 12). This indicates that the cognitive load (Sweller, 1988) of validating these multi-source, AI-integrated solutions often outstrip the user's capacity, leading to a reliance on trust despite declining performance.

## 5.4. Context-dependent ethics and domain-specific trust

The Wave 3 data robustly confirm the proceeding of context-dependent ethical judgment. The clear ethical discrimination observed in earlier waves remains stark, as a majority approved of AI use in private (65.2% "not cheating") and professional (69.6% "not cheating") contexts, while an identical majority (69.6%) condemned its use in academic theses as "cheating" (Table 3). This stability reinforces the conclusion that ethical calibration is a fundamental aspect of user behavior, aligning with framework approaches that emphasize domain-specific norms (Floridi & Cowls, 2022).

A refinement from Wave 3 lies in the interpretation of trust metrics. The observed pattern (i.e. moderate reliability (M=4.9/8) and data safety (M=4.9/8) assessments co-occurring with universal, daily use) now appears not just as pragmatic adoption, but as a mature, calibrated trust (Glikson & Woolley, 2020). Users are not naive; they are aware of the technology's limitations. Their high usage reflects a calculated risk-benefit analysis where the utility for professional (91.3%) and educational (82.6%) tasks (Table 2) outweighs the moderate perceived risks. This suggests that trust is not a prerequisite for use, but an evolving judgment formed through repeated, context-specific experience.

### 5.5. Theoretical Integration: Toward a Multidimensional Framework toward a human-AI collaboration

Integrating the empirical confirmations and new patterns from Wave 3, we can strengthen the Model of Calibrated Human-AI Collaboration in (Huemmer et al., 2025a), (Huemmer et al., 2025b) and (Huemmer et al., 2025c) into a more robust theoretical framework. The four interdependent dimensions are updated as follows:

1. Strategic Integration: This dimension is confirmed as the foundational layer. The prevalence of human-first workflows (Table 5) underscores that effective collaboration is orchestrated, not delegated. This extends distributed cognition theory (Hutchins, 1995) by highlighting that the human role is evolving from a direct solver to a cognitive workflow manager.
2. Verification Self-Efficacy: Wave 3 data elevate the criticality of this dimension. The pronounced proof-belief gap (Table 14) demonstrates that verification self-efficacy is not merely a mediator but the primary bottleneck for effective collaboration on complex tasks. This concept must be central to extending technology acceptance models (Venkatesh et al., 2003), as perceived usefulness is moot if users cannot validate the outputs they receive.
3. Context-Dependent Appropriateness: The stability of ethical judgments (Table 3) confirms this as a core, stable dimension of the collaboration. It moves from an observation to a principle: effective and ethical AI use is inherently situated. Frameworks must account for the distinct normative landscapes of professional, private, and academic domains (Floridi & Cowls, 2022).
4. Metacognitive Calibration: The quantification of the belief-performance gap (Table 14) provides a crucial update to this dimension. It is no longer sufficient to appraise one's own knowledge; users must also develop accuracy calibration regarding the AI's output. This involves recognizing that fluent, confident AI responses can be profoundly incorrect, especially in complex domains, and adjusting trust and verification effort accordingly (Ji et al., 2023).

This synthesized model, updated with the Wave 3 findings, provides a comprehensive framework for diagnosing challenges and designing interventions in human-AI collaboration, specifying that successful outcomes depend on managing workflows, building verification skills, respecting contextual norms, and maintaining clear-eyed accuracy calibration.

### 5.6. Connection of Wave 3 data to Wave 1 and 2 Study Results and Theoretical Synthesis

The findings from Wave 3 (n=23) robustly confirm, refine, and extend the core phenomena identified in the initial waves (Huemmer et al., 2025b, c), solidifying the empirical foundation for understanding human-AI collaboration in problem-solving.

#### 5.6.1. Confirmation and Refinement of Core Phenomena

Wave 3 data strongly reinforce the necessity of the AI supervision skills framework. The consistent observation of hybrid workflows beginning with "Think" (e.g., "Think, Internet, ChatGPT, Further Processing" at 39.1% in Table 5) validates Prompt Framing and Decomposition as a critical competency, confirming the pattern from earlier waves that effective sequences are predicated on initial cognitive structuring.

Furthermore, the continued high reliance on Strategic Triangulation is confirmed. The pervasive use of the Internet (95.7%) and consultation with peers (78.3%) (Table 6) demonstrates that users instinctively cross-verify AI outputs against other sources. This refinement (showing that social consultation, while high, is not universal) provides a more nuanced picture of triangulation behaviors, aligning with practices for mitigating cognitive biases (Nickerson, 1998; Sullivan et al., 2023).

The central concept of a verification paradox is powerfully corroborated. Wave 3 shows that AI consultation is highest for difficult problems (73.9%, Table 8) and complex vignettes (e.g., 69.6% for Problem 3, Table 11), while the self-assessed ability to prove correctness for complex problems drops to 59.1% (Table 10). This creates a critical vulnerability, as seen in the significant drop in objective performance on the most complex task (13.0% for Problem 4, Table 13). This pattern of automation bias, where users over-rely on automated aids (Parasuraman & Manzey, 2010), is critically exacerbated in LLMs due to their fluent and opaque outputs (Singh et al., 2025).

#### 5.6.2. Theoretical Integration: Toward a Multidimensional Model of Calibrated Collaboration

Integrating the empirical evidence from all three waves, we now present a refined Model of Calibrated Human-AI Collaboration initially proposed in Huemmer et al. (2025b), the Wave 3 data solidify the four interdependent dimensions:

1. Strategic Integration: Confirmed by the persistent dominance of human-led, hybrid workflows (Table 5). This dimension, grounded in distributed cognition (Hutchins, 1995), positions the human as the essential orchestrator of the problem-solving process.
2. Verification Self-Efficacy: Wave 3 confirms this as the critical bottleneck. The data in Table 10 shows a clear decline in the perceived ability to verify solutions as complexity increases (from 86.4% to 59.1%), extending technology acceptance models (Venkatesh et al., 2003) by identifying verification confidence as a crucial mediator between use and effective outcomes.
3. Context-Dependent Appropriateness: The stability of this dimension is striking. Wave 3 data (Table 3) robustly confirm the pattern of high acceptance in private (65.2%) and professional (69.6%) contexts, contrasted with strong reservations in academic settings (69.6% cheating). This demands nuanced, domain-aware policies (Floridi & Cowls, 2022).
4. Metacognitive Calibration: Wave 3 allows for the quantification of this dimension through the belief-performance gap. The stark contrast, especially in Problem 4 where perceived correctness was 93.8% but actual correctness was 13.0% (Table 14), reveals a profound miscalibration. This integrates research on metacognitive monitoring, emphasizing the need to accurately judge both personal capabilities and AI limitations (Dunning et al., 2004; Fan et al., 2024; Kruger & Dunning, 1999; Lee et al., 2025).

In conclusion, Wave 3 does not merely replicate earlier findings; it provides richer, more granular data that solidifies the observed patterns of hybrid problem-solving and clarifies the verification bottleneck as the most pressing vulnerability in current human-AI collaboration.

#### 5.6.3. Practical and Educational Implications: From Identification to Implementation

The convergence of findings across waves provides compelling evidence for targeted interventions and underscores the urgency of implementing the following proposed practical and educational scaffolds. First, verification scaffolds are necessitated by the performance data documented in Table 13. These must include embedded cognitive supports such as Assumption Documentation Protocols to counter foundational logic neglect (Chen et al., 2024), Triangulation Procedures to institutionalize multi-source verification (Sullivan et al., 2023), and Adequacy Criteria Checklists to operationalize abstract validation processes (Kasneci et al., 2023).

Second, educational imperatives emerge from the concentration of AI use in high-cognitive domains, specifically research (82.6%) and verification (69.6%, Table 7). This pattern mandates curricula focused on critical AI literacy that transcends operational competence (Ng et al., 2024). Essential competencies include decompositional thinking for prompt optimization (Zhang et al., 2024), hypothetico-deductive verification methodologies (Sullivan et al., 2023), systematic assumption tracking (Chen et al., 2024), and failure mode recognition (Ji et al., 2023).

### 5.7. Practical implications: Scaffolding Verification Processes

The belief-performance and proof-belief gaps identified and quantified across all three waves of this longitudinal study highlight verification as the critical bottleneck in complex AI-assisted problem-solving (Huemmer et al., 2025a, 2025b, 2025c). The quantification of belief-performance and proof-belief gaps across all waves (Table 14) identifies verification as the fundamental constraint in complex AI-assisted problem-solving. To address this systemic challenge, we propose evidence-based verification scaffolds that structured supports designed to mitigate identified cognitive vulnerabilities.

The first scaffold involves Assumption Documentation Protocols, which mandate the explicit articulation of premises, constraints, and domain assumptions prior to AI consultation. This intervention directly counteracts the observed tendency to accept AI outputs without evaluating their foundational logic, a key contributor to the belief-performance gap (Chen et al., 2024). A second scaffold is the implementation of Adequacy Criteria Checklists, which provide domain-specific, stepwise frameworks for evaluating AI-generated solutions. By operationalizing the abstract process of "verification," these checklists reduce cognitive load (Sweller, 1988) and help bridge the proof-belief gap by making validation processes both explicit and manageable (Kasneci et al., 2023). Finally, Triangulation Procedures institutionalize cross-verification across multiple independent sources, methodologies, or AI architectures. This scaffold formalizes the healthy skepticism noted in user behaviors and directly counters automation bias by preventing over-reliance on a single source (Sullivan et al., 2023).

The implementation of such scaffolds is a direct, practical response to the paradigm of orchestrated integration observed in this study, ensuring that the human role as a "solution validator" is effectively supported rather than overwhelmed.

### 5.8. Educational implications: cultivating critical AI literacy

The empirical evidence from this longitudinal study necessitates a paradigm shift in educational approaches toward developing critical AI literacy. The escalating verification gaps with task complexity (Table 14) provide unequivocal evidence for educational frameworks that integrate technical proficiency with epistemic vigilance (Kasneci et al., 2023; Singh et al., 2025).

The concentration of AI application in high-level cognitive domains, particularly research (82.6%) and correctness verification (69.6%, Table 7), clearly delineates the precise contexts where future educational interventions will yield the greatest impact. This finding necessitates a move beyond basic operational competence toward a form of critical AI literacy that encompasses the specific epistemic practices essential for rigorous human-AI collaboration (Ng et al., 2024).

A set of core competencies, derived from patterns observed across the study waves, forms the foundation of this literacy. These include Decompositional Thinking, which involves training in problem framing and subsystem identification prior to AI engagement to enhance input quality and output reliability (Zhang et al., 2024). A second competency is Hypothetico-Deductive Verification, which focuses on developing methodologies to treat AI outputs as provisional hypotheses requiring systematic testing through counterexample generation and boundary condition analysis (Sullivan et al., 2023). Furthermore, Assumption Tracking entails implementing explicit protocols for documenting premises and constraints throughout the problem-solving lifecycle, thereby making implicit reasoning explicit and evaluable (Chen et al., 2024). Finally, Failure Mode Recognition involves cultivating an anticipatory awareness of characteristic AI failure patterns, such as hallucination, context collapse, and reasoning shortcuts (Ji et al., 2023).

These competencies are optimally developed through targeted pedagogical approaches. These include the analysis of examples that incorporate expert verification traces, the use of rubric-based adequacy criteria aligned with disciplinary standards, and the implementation of structured peer-review mechanisms that focus specifically on validation processes rather than solely on final outputs (Kasneci et al., 2023).

## 6. Limitations and Future research

The empirical findings from Waves 1, 2, and 3 of this longitudinal investigation must be interpreted within well-defined methodological and conceptual constraints that simultaneously delineate productive avenues for subsequent research (Huemmer et al., 2025a, 2025b, 2025c).

### 6.1. Methodological Constraints and Boundary Conditions

Several methodological characteristics establish important boundary conditions for interpreting the observed patterns. First, the reliance on self-report measures for critical constructs, which that including perceived correctness (Table 12), verification capacity (Table 10), and ethical judgments (Table 3), introduces potential common method variance that may systematically influence the precise quantification of belief–performance and proof–belief gaps (Podsakoff et al., 2003; 2012). This limitation is particularly relevant given the documented discrepancies between self-assessed and demonstrated capabilities across complexity levels.

Second, while the problem vignettes were theoretically grounded in cognitive load frameworks (Sweller, 1988) and followed established complexity progressions (Huemmer et al., 2025a), they lacked formal psychometric calibration. The dramatic performance differential between Problem 3 (66.7% accuracy) and Problem 4 (13.0% accuracy, Table 13) raises the possibility that observed effects reflect item-specific features rather than pure complexity gradients (Wilson, 2023). This limitation necessitates caution in generalizing the specific performance patterns beyond the current task domain.

Finally, the sampling framework that is characterized by high baseline AI literacy (95.7%, Table 2) and recruitment from academic technology communities establishes a clear boundary for generalizability. The observed verification behaviors and ethical calibrations (Table 3) likely represent optimal-case scenarios that may not extend to populations with lower technological fluency or different institutional affiliations (Huemmer et al., 2025c).

### 6.2. Empirical and Theoretical Research Trajectories

The methodological constraints identified in Study Wave 3 directly inform a structured research agenda designed to address both empirical and theoretical priorities. First, a program of psychometric refinement is necessary, involving the development and validation of a calibrated task bank with established difficulty parameters across multiple problem domains. This would enable a more precise disentanglement of complexity effects from item-specific variance, building on modern measurement frameworks (Wilson, 2023).

Second, the agenda calls for a shift towards direct behavioral measurement to move beyond reliance on self-report data. Implementing technologies such as source click-through rate analysis, eye-tracking during solution evaluation, and process-tracing methodologies would provide objective validation of the observed gaps between belief in a solution and the capability to verify it (Zhou et al., 2024).

A third priority involves conducting experimental interventions to establish causal efficacy. This entails the controlled manipulation of verification scaffold conditions, such as mandatory assumption documentation, adequacy checklists, and triangulation requirements, to test their power in mitigating the identified cognitive gaps (Chen et al., 2024; Sullivan et al., 2023).

Finally, a longitudinal extension of the multi-wave design is essential for tracking the evolutionary trajectory of AI-assisted problem-solving architectures. Continuing this temporal analysis is critical for understanding the durability of intervention effects and the transfer of critical AI literacy competencies across different domains (Huemmer et al., 2025c). Collectively, this research agenda directly addresses the methodological constraints identified in the present study while simultaneously advancing the theoretical understanding of the cognitive architectures that underline effective human-AI collaboration.

## 7. Conclusion

The investigation in Wave 3 of this multinational longitudinal study definitively characterizes a paradigm of orchestrated human-AI collaboration in problem-solving, confirming a model of strategic integration over indiscriminate reliance (Huemmer et al., 2025a, 2025b). This collaborative framework, however, is fundamentally constrained by a critical and escalating verification deficit that constitutes the primary vulnerability within this emerging cognitive ecosystem. Through longitudinal behavioral tracking, this research has empirically identified and precisely quantified two systematic epistemic gaps, namely a belief-performance gap, where subjective confidence in AI-generated solutions substantially exceeds objective accuracy on complex tasks (reaching +80.8 percentage points for Problem 4, Table 14), and a proof-belief gap, where demonstrable verification capability consistently lags behind perceived correctness (reaching -16.8 percentage points for Problem 3, Table 14). These findings, contextualized within the theoretical framework of cognitive redistribution and calibrated trust (Glikson & Woolley, 2020), establish that the fundamental constraint on reliable AI-assisted work is not solution generation but solution validation.

The primary practical implication of these Wave 3 findings necessitates a paradigm shift in both technological design and educational focus from optimizing for fluent generation to scaffolding robust verification processes. The empirical evidence for a consolidated hybrid problem-solving culture, evidenced by the predominance of structured workflows (Table 5) and strategic triangulation behaviors (Table 6), underscores that the most critical human contributions will increasingly center on cognitive orchestration, critical evaluation, and ethical governance. Consequently, the ultimate efficacy of human-AI problem-solving will depend on our collective capacity to embed powerful generative tools within workflows and learning environments that render verification processes explicit, pedagogically tractable, and systematically accountable, as operationalized in the proposed scaffolds and educational imperatives (Chen et al., 2024; Kasneci et al., 2023). The findings from Wave 3 provide a definitive empirical foundation for this necessary evolution, highlighting the urgent imperative to fortify the human role as a critical validator in the age of generative AI.

## 9. Appendix – Answers from Participants

**Table 15.** Self-reported process steps to solve a difficult or complex problem (i.e. answer Others selected).

| **Answers about the own approach to solve a difficult problem** |
| --- |
| I cannot answer this question |
| Think, ask others, internet, further processing, chatgpt |
| Think, Paper, Sketch, Internet Researche - What suites my problem best? - Book, Pure Internet of ChatGPT, mostly a combination of two , Further Processcing |
| Think Paper Sketch Internet Paper ChatGPT Further Processing |

**Note.** Original Answers and hence not corrected.

**Source.** Authors' own illustration based on survey data.

**Table 16.** Self-reported problem-solving approach for a *simple* problem.

| **Answers about the own approach to solve a simple problem** |
|---|
| I first try to solve it my own but if I fails I take help in some steps |
| SKETCH, THINK, PAPER, REFORMULATION |
| solve by my own knowledge or check the book if I had a problem to remind some parts. And if it was not serious just search in internet |
| Think> Internet > Chat GPT > Further processing |
| Think, Further Processing, ChatGPT |
| think, further processing. |
| Think, Internet, ChatGPT and book |
| Think, sketch, think further |
| THINK, SOLVE, INTERNET, REFORM |
| Think, writing and organize ideas, research on internet, and chat gpt |
| Think about the problem, visualize it on paper, work through any remaining steps |
| Think and as ChatGPT |
| think and calculate |
| Think and determine |
| Think and secure with chatgpt or go straight to chatgpt |
| think and sketch |
| Think Solve |
| Tink and then solve |

**Note.** Original Answers and hence not corrected.

**Source.** Authors' own illustration based on survey data.

**Table 17.** Self-reported problem-solving approach for a *difficult* problem.

| **Answers about the own approach to solve a difficult problem** |
|---|
| Ask ChatGPT |
| chatgpt, check via google, maybe reformulate -> chatgpt |
| Check the book, ask AI (like chatgpt), Check Chatgpt answers, if it was not good search internets for other resources |
| I first try it on my own take help from Internet and also from chatgpt |
| Internet research, youtube videos, and chat gpt |
| Think > AI (ChatGPT) > Further Processing |
| think, calculate or sketch (chatGPT only if needed) |
| think, chatgpt, further processing |
| Think, ChatGPT, Reformulation |
| Think, Internet, ChatGPT, book, asking others |
| think, internet, own approach |
| THINK, INTERNET, REFORM |
| THINK, INTERNET, REFORMULATION |
| Think, sketch, internet |
| Think about the problem, visualize it on paper, work through any remaining steps |
| Think and internet |
| Think Research if required Solve Check |
| Think sketch and solve |

**Note.** Original Answers and hence not corrected.

**Source.** Authors’ own illustration based on survey data.

**Table 18.** Self-reported problem-solving approach for a *complex* problem.

| **Answers about the own approach to solve a complex problem** |
| --- |
| Books, Papers, Youtube or other course, Internet |
| chatgpt, check via google, maybe reformulate -> chatgpt |
| I do not have complex problems |
| research internet or library, think, pair the results with genAI, reflect |
| Take help from books from Internet |
| Think > AI > Further processing |
| think, calculate (if it is mathematical problems) or internet or chatGPT, further processing |
| think, chat gpt, further processing |
| THINK, INTERNET, AI, REFORM |
| Think, internet, chatgpt |
| Think, Internet, ChatGPT, book and others |
| THINK, INTERNET, REFORMULATION |
| Think about the problem, visualize it on paper, work through any remaining steps |
| Think and internet |
| Think Research ChatGPT Check reformulation if required ChatGPT Check |
| Think Sketch AI and then solve |
| writing down what I need to research, youtube videos for possible, asking for reliable sources on chat gpt, and the going to the different links provided |

**Note.** Original Answers and hence not corrected.

**Source.** Authors' own illustration based on survey data.